\documentstyle[aps,multicol,psfig]{revtex}

\begin{document}
\newcommand{\bra}{\langle}
\newcommand{\ket}{\rangle}
\bibliographystyle{prsty}
\widetext
\title{
Spectral correlations: understanding
oscillatory contributions}
\author{B. Mehlig$^1$ and M. Wilkinson$^2$}
\address{\mbox{}$^1$Max-Planck-Institut f\"ur Physik
           komplexer Systeme, N\"othnitzer Str. 38, 01187 Dresden, Germany\\
         \mbox{}$^2$Department of Physics and Applied Physics,
           University of Strathclyde, Glasgow, Scotland, UK}
\date{\today}
\maketitle{} 
\begin{abstract}  
We give a transparent derivation of a relation 
obtained using a supersymmetric non-linear sigma model
by Andreev and Altshuler 
[{\sl Phys. Rev. Lett.} {\bf 72}, 902, (1995)], which connects smooth and 
oscillatory components of spectral correlation functions.
We show that their result is not specific 
to random matrix theory. Also, we show that despite an apparent
contradiction, the results obtained
using their formula are consistent with earlier perspectives on 
random matrix models. In particular,
the concept of resurgence is not required. 
\end{abstract}   
\begin{multicols}{2}
Spectral correlations of complex quantum systems, such as 
disordered metals and classically chaotic quantum systems, 
are known to be nearly universal. For small ranges of energy 
they are well approximated 
by the Gaussian invariant ensembles of random matrix theory 
(GXE, where X=O,U or S stands for orthogonal, unitary and 
symplectic invariance) \cite{haa91,meh91}.
Deviations from GXE behaviour at larger
energy scales may be consistently incorporated
using semiclassical or perturbative approaches \cite{ber85,wil88,alt79}.
An interesting development in this field
was a paper by Andreev and Altshuler (AA) \cite{and95}, who
introduced a relation which suggests a degree
of non-universality in short ranged spectral 
correlations. Their calculations are
based on the non-linear sigma model.
Our paper will give a transparent physical
insight into their relation.

AA \cite{and95} consider the spectral
two-point correlation function, defined as
\begin{equation}
\label{eq:1}
R_\beta (\epsilon) = \Delta^2\, 
\langle d(E+\epsilon/2)\, d(E-\epsilon/2)\rangle -1\,.
\end{equation}
Here  $d(E) = \sum_n \delta(E-E_n)$ is the density
of states, $E_n$ are the eigenvalues of
a Hamiltonian $\widehat H$ and
$\Delta(E) = \langle d(E)\rangle^{-1}$ is the
mean level spacing which is assumed to be independent
of $E$ in the following. The index $\beta=1,2,4$
distinguishes orthogonal, unitary or symplectic symmetry 
classes respectively.

AA divide 
Eq. (\ref{eq:1}) into a smooth and an oscillatory
contribution, and propose that (for $\beta=2$) these are related
as follows
\begin{eqnarray}
\label{eq:2}
R_2^{\rm av}(\epsilon) 
&\simeq& -\frac{\Delta^2}{4\pi^2}\frac{\partial^2}{\partial \epsilon^2}\,
\log\,D(\epsilon)\,\\
\label{eq:3}
R_2^{\rm osc}(\epsilon) &\simeq&\frac{1}{2\pi^2}\,
\cos(2\pi\epsilon/\Delta)\, D(\epsilon)\,.
\end{eqnarray}
The relation is exact for the GUE with a suitable choice for $D(\epsilon)$
and generalizations are approximately true for the other 
ensembles.
AA propose that Eqs. (2) and (3) are quite general, and should be used
to predict the oscillatory component from the average
component. The quantity $D(\epsilon)$ is a spectral
determinant.

The AA relation contained in (\ref{eq:2}) and (\ref{eq:3}) has attracted 
considerable attention (see, for example, 
\cite{aga95,bog96,pra96}), 
partly because of the suggestion that it contains
information about \lq resurgence'. This is to be 
interpreted in terms of Gutzwiller's \cite{gut90} relationship
between periodic orbits and the density of states: 
periodic orbits orbits with period $t_j$ are associated
with oscillations in the density of states of period
$\epsilon_j=\hbar/t_j$. The concept of resurgence is
that information about long period orbits is encoded in
properties of the short orbits. In this context, 
$R^{\rm av}(\epsilon)$ may be derived using Gutzwiller's
relation from properties of short orbits, and the AA
relation then gives information about fluctuations in the spectrum
with wavelength equal to the mean level spacing $\Delta$,
corresponding to orbits of period equal to the 
\lq Heisenberg time', $t_{\rm H}=2\pi\hbar/\Delta$.

This interpretation of the AA result is at variance
with some old ideas about the applicability of 
random matrix theory. Dyson \cite{dys62} introduced 
a Brownian motion model for random matrix
theory, giving a Langevin equation of motion for the response
of energy levels to a stochastic perturbation. Dyson later suggested
\cite{dys72} that this model may give valuable insights into why random
matrix theory applies to generic systems. He considered
the dynamics of the energy levels under the effect of a 
stochastic perturbation, which is large enough to shift
energy levels significantly, but which remains small enough  
to leave other features of the system (such as the classical 
dynamics) unchanged. Dyson showed that
the equations of motion simplify if the discrete Fourier 
coefficients $a_k$ of the level displacements are used 
as dynamical variables:
\begin{equation}
\label{eq:4}
a_k = \frac{1}{N} \sum_{n=0}^{N-1} \Delta E_n \,\exp(-2\pi {\rm i} kn/N)
\end{equation}
where $\Delta E_n = E_n-n\Delta$, $N$ is the number of energy
levels, and $k$ takes $N$ successive integer values. We will
take the maximal $k$ to be ${\rm int}(N/2)$. The long wavelength modes
evolve almost independently, with a long
relaxation time, which scales as $k^{-1}$ \cite{dys72,wil88}.
The short wavelength modes remain strongly coupled. 
The stochastic perturbation will bring the short 
wavelength modes into equilibrium, giving statistics
of the $a_k$ which are identical to the random matrix ensemble.

These arguments were later supported and extended
with the aid of semiclassical estimates of matrix elements 
\cite{wil88}. The argument in \cite{dys72} assumes that
the matrix elements of the perturbation are uncorrelated.
It was shown that semiclassical estimates are consistent with this
hypothesis when considering the stochastic force driving the
large $k$ modes. 
The forces driving the small $k$ modes
are modified by the classical dynamics of the system. 
The resulting picture is that long wavelength fluctuations
are non-universal, but that at short wavelengths
the excitations of the modes are precisely the same
as for the appropriate GXE. In particular, there is no 
modification of the statistics of the Fourier coefficients $a_k$
unless $\vert k\vert /N$ is small. This appears to be
in contradiction with the AA relations, which suggest that
non-universal corrections to the smooth part of the 
correlation function are echoed by oscillations
with large wavenumbers $k$.

One resolution of this contradiction would be that there
are previously un-suspected correlations between
matrix elements due to \lq resurgence', which are not
captured by the semiclassical approximations in \cite{wil88}. 
We will however show that there is no need in invoke 
this principle. We will first describe a simple derivation of the AA
result, and comment on its applicability. We will then
describe the Brownian motion model and use it to derive 
the correlation function $R_\beta(\epsilon)$ for the 
case of a system with diffusive electron motion. 
In this calculation we assume that the statistics of the
Fourier coefficients are unchanged for large $\vert k\vert$.
The fact that we reproduce existing results verifies
that resurgence is not an essential ingredient in their
derivation. Finally we will comment on the correlation
function for classically chaotic systems. 

Our starting point is the following
general expression for the correlation
function $R_\beta(\epsilon)$:
\begin{equation}
\label{eq:5}
R_\beta(\epsilon) = -1+\Delta
\sum_{n=-\infty}^\infty p_\beta(n,\epsilon)
\end{equation}
where $p_\beta(n,\epsilon) {\rm d}\epsilon$
denotes the probability of finding
that the difference between $E_0$ and $E_{n}$
is in the interval $[\epsilon,\epsilon+{\rm d}\epsilon]$.
We will show below that in many cases,
the $p_\beta(n,\epsilon)$ are
well approximated by Gaussians for large $n$.
We will estimate the variance $\sigma^2_\beta(n)$ for 
different systems.

First, however,
we show how the relations (\ref{eq:2}) and (\ref{eq:3})
can be derived from (\ref{eq:5}).
We write
\begin{equation}
\label{eq:6}
R_\beta(\epsilon) = R_\beta^{\rm av}(\epsilon)+ R_\beta^{\rm osc}(\epsilon)
\,.
\end{equation}
Here $R_\beta^{\rm av}(\epsilon)$ is defined as
\begin{equation}
\label{eq:7}
R_\beta^{\rm av}(\epsilon)
 = \int_{-\infty}^\infty \!{\rm d}\epsilon^\prime\,
w(\epsilon-\epsilon^\prime)\,R_\beta(\epsilon')
\end{equation}
where $w(\epsilon)$ is a suitable window function (which
could be a Gaussian centred around zero 
with variance much larger than $\sigma^2_\beta(n)$ and normalized
to $\Delta^{-1}$). 
$R_\beta^{\rm osc}(\epsilon)$ is the remaining
oscillatory contribution.
Using (\ref{eq:5}) we have,
upon expanding the slowly varying window function in (\ref{eq:7})
\begin{eqnarray}
\label{eq:8}
R_\beta^{\rm av}(\epsilon) 
 &=&-1+\Delta\!\!\!\sum_{n=-\infty}^\infty
\int_{-\infty}^\infty 
\!{\rm d}\epsilon'\, w(\epsilon-\epsilon')\,p_\beta(n,\epsilon')\\
&\simeq& 
-1+\Delta\!\!\!\sum_{n=-\infty}^\infty \!\!
\big[w(\epsilon\!-\!n\Delta) +\frac{1}{2} w''(\epsilon\!-\!n\Delta)
\sigma_\beta^2(n)\big]  \,.
\nonumber
\end{eqnarray}
Assuming that $\sigma^2_\beta(n)$ is a slowly varying function of $n$,
we may approximate the sum over $n$ as an integral and obtain
\begin{equation}
\label{eq:9}
R_\beta^{\rm av}(\epsilon) \simeq \frac{1}{2\Delta^2} 
\frac{\partial^2}{\partial n^2}\,\sigma_\beta^2(n) \,,
\hspace*{0.2cm} \epsilon = n\Delta\,.
\end{equation}
Consider now the oscillatory contribution 
$R_\beta^{\rm osc}(\epsilon)$. Again assuming that
$\sigma^2_\beta(n)$ is slowly varying, these can
be evaluated from Eqs. (\ref{eq:5}) 
using Poisson summation:
\begin{equation}
\label{eq:10}
R_\beta^{\rm osc}(\epsilon)
\simeq
\sum_{\mu = 1}^\infty 
2\cos( 2\pi\,\mu n)\,\,\,
{\rm e}^{\displaystyle -(2\pi\mu)^2\sigma^2_\beta(n)/2\Delta^2}
\,.
\end{equation}
This sum is dominated by the $\mu=1$ term.
Defining 
$D(\epsilon) = 
4\pi^2 \exp[-2\pi^2\sigma^2(n)/\Delta^2]$
we obtain the relations
(\ref{eq:2}) and (\ref{eq:3}).
We remark that these formulae are valid
for any correlation function that can
be approximated as a sum of Gaussians, in regions where
the variance satisfies $\sigma^2\gg \Delta^2$, and where
$\sigma^2$ varies sufficiently slowly as a function of $n$.
Relating $D(\epsilon)$ to $\sigma^2 (n)$ gives
a clear insight into its meaning. 
The relations 
(\ref{eq:2}) and (\ref{eq:3}) may be
easily extended by considering higher-order
derivatives in (\ref{eq:2}) and
higher Fourier components in (\ref{eq:3}).
Extensions to non Gaussian spacing distributions
are possible.
Also, we emphasize that these relations
are not specific to spectral correlation functions.
For example, they are applicable to density
correlations in solids: for thermal excitation
of phonons the corresponding
two-point function is a sum of Gaussians, and 
$\sigma^2(n)\sim \log n$ in $d=2$ dimensions \cite{pei79}.

Next we consider the calculation of 
$p_\beta(n,\epsilon)$ and $\sigma_\beta^2(n)$
using Dyson's Brownian motion model.
In the Brownian motion model, the matrix elements of
the Hamiltonian $\widehat H$ undergo a diffusive evolution
as a function of a fictitious time variable
$\tau$. We denote the infinitesimal change
of $\widehat H$ by $\delta \widehat H$.
In the GOE case, we have for $n>m$ and $n'>m'$,
\begin{eqnarray}
\label{eq:11}
&&\langle\delta H_{mn}\rangle = 0\,,\hspace*{0.2cm} 
\langle \delta H_{mn} \,\delta H_{m'n'}\rangle = 
C_{mn}^{\rm off}\,\delta\tau\, \delta_{nn'} \,\delta_{mm'}
\end{eqnarray}
(extensions for GUE and GSE are given in \cite{dys62}). 
The diagonal elements obey
\begin{equation}
\label{eq:12}
\mbox{}\hspace*{-5mm}
\langle \delta H_{nn}\rangle = 0\,,\hspace*{0.2cm}
\langle \delta H_{mm}\, \delta H_{nn}\rangle
 = 2\,\delta\tau\,\beta^{-1}\,C_{mn}^{\rm diag}\,.
\end{equation}
Dyson \cite{dys62} originally discussed the case where 
$C_{mn}^{\rm diag} = \delta_{mn}$ and $C_{mn}^{\rm off} = 1$,
for which the statistics of the $E_n$ are the same as for
the GXE.
We will argue below that non-universal
deviations from the
 GXE are encoded in $C_{mn}^{\rm diag} = C_{m-n}^{\rm diag}$
(and $C_{mn}^{\rm off} = C_{m-n}^{\rm off}$).
Using second order perturbation theory leads to a Langevin equation
\begin{equation}
\label{eq:13}
\delta E_n = \sum_{m\neq n} \frac{|\delta H_{mn}|^2}{E_n-E_m}
+\delta H_{nn}
\end{equation}
for the energy level shifts $\delta E_n$.  Thus
\begin{eqnarray}
\label{eq:14}
&&\langle \delta E_n \rangle = \delta\tau\sum_{m\neq n} 
\frac{C_{m-n}^{\rm off}}{E_m-E_n}\,,\\
\label{eq:15}
&&\langle\delta E_m\delta E_n \rangle =  2\,\delta \tau\,\beta^{-1}\,
C_{m-n}^{\rm diag}\,.
\end{eqnarray}
Semiclassical estimates
\cite{wil87} indicate that the $C^{\rm off}_n$ decrease
for large values of $n$, i.e. the repulsive interaction is screened
at long range. This effect was considered in \cite{cha96};
for our purposes it is not significant, and we will set $C^{\rm off}_n=1$
throughout.
We now use (\ref{eq:4}) to obtain the same equations of motion in terms
of the Fourier variables $a_k$. Using $\delta\Delta E_n = \delta E_n$, 
we have
\begin{equation}
\label{eq:16}
\langle \delta a_k \delta a_l^\ast \rangle =
2\,\delta\tau\,\beta^{-1}\,I_k\,\delta_{kl}
\end{equation}
where $I_k = N^{-1}\sum_{n} C_{n}^{\rm diag} \exp(-2\pi {\rm i} kn/N)$.
The transformation of the drift term is less straightforward.
In general the expectation value of $\delta a_k$ is a complicated
function of all of the $a_k$, but in the limit $\vert k\vert /N\to 0$ the
equations decouple and obey \cite{wil88,dys62}
\begin{equation}
\label{eq:17}
\langle \delta a_k\rangle =  -{2\pi^2\,k\over{N\Delta^2}}a_k\, \delta\tau\,.
\end{equation}
By solving a Fokker-Planck equation for the real and imaginary
parts of $a_k$, these are found to have a steady state distribution 
which is Gaussian. The value of $\langle \vert\ a_k \vert^2\rangle$ 
can be deduced by requiring that 
$\langle \vert a_k(\tau+\delta \tau)\vert^2\rangle
=\langle \vert a_k(\tau)\vert^2\rangle$: writing 
$a_k(\tau+\delta \tau)=a_k(\tau)+\delta a_k$ and 
using (\ref{eq:16}) and (\ref{eq:17}) we find, 
provided $\vert k\vert /N\ll 1$:
\begin{equation}
\label{eq:18}
\langle |a_k|^2\rangle =  \frac{N\Delta^2 I_k}{2\pi^2 \beta k}\,.
\end{equation}
We cannot determine $\langle \vert a_k \vert^2 \rangle$ 
for larger values of $\vert k\vert $ from this approach. 
We now make our key assumption, that $I_k\sim N^{-1}$ 
when $\vert k\vert /N$ is 
not small. This corresponds to assuming that there are no short-ranged 
correlations between the diagonal matrix elements, i.e. 
$C^{\rm diag}_{n-m}\sim \delta_{nm}$.  Semiclassical arguments
which support, but do not prove, this assumption are given in \cite{wil88}.
(The existence of short ranged correlations would represent 
a type of \lq resurgence', in the sense discussed earlier). 
Under this assumption, 
the equations of motion of the $a_k$ are identical to those for
the Brownian motion model describing Gaussian invariant ensembles.
We therefore conclude that when $\vert k\vert /N$ is not small, the mode 
intensities $\langle \vert a_k \vert^2 \rangle$ are identical
to those of the Gaussian invariant ensembles.

For large $n$, the level spacings $E_n-E_0$ are seen
to be a sum of many independent random variables
and are thus Gaussian distributed
\begin{equation}
\label{eq:19}
p_\beta(n,\epsilon) 
= [2\pi\sigma_\beta^2(n)]^{-1/2}\,
{\rm e}^{\displaystyle -(\epsilon-n\Delta)^2/2\sigma^2_\beta(n)}
\end{equation}
with mean $n\Delta$ and with variance
\begin{equation}
\label{eq:20}
\sigma^2_\beta(n) = 4\sum_k \langle |a_k|^2\rangle\, 
\sin^2\big(\frac{\pi  n k}{N}\big)\,.
\end{equation}
Writing $t=t_{\rm H}k/N$, and $NI_k=I(t)$, Eq. (\ref{eq:20}) becomes
\begin{equation}
\label{eq:21}
\sigma^2_\beta(n) = \frac{4\Delta^2}{\beta\pi^2}
\int_0^{t_{\rm H}/2}\!\frac{{\rm d}t}{t}\,J_\beta(t)\,
\sin^2(\frac{n t\Delta}{2\hbar})
\end{equation}
where $J_\beta(t)=I(t)$ for $t\ll 1$, and $J_\beta(t)$
takes a universal (but to us, unknown) form when $t$ is not small.
For the Gaussian invariant ensembles (where $I(t)=1$), the variances 
clearly grow logarithmically with $n$. They are, asymptotically
for large $n$,
\begin{equation}
\label{eq:22}
\sigma^2_\beta(n) \simeq \frac{2\Delta^2}{\beta\pi^2} 
\left[\log(2\pi n) + C_\beta\right]
\end{equation}
with $C_1 = -\log\sqrt{2}$, $C_2 = 0$ and $C_4=\log(4/\pi)$.
Using this expression in (\ref{eq:2}) and (\ref{eq:3}) gives
the correct leading order contributions to 
$R_\beta^{\rm av}(\epsilon)$ and $R_\beta^{\rm osc}(\epsilon)$
in the limit $\epsilon/\Delta \to \infty$ (c.f. \cite{meh91}).

Next we consider how the function $I(t)$ must be modified
at small $t$ to take account of classical dynamics.
For small values of $\vert k\vert $, the amplitude $\delta a_k$ can
be estimated semiclassically \cite{wil88}
\begin{equation}
\label{eq:23}
\delta a_k \sim {1\over N}{\rm tr}[\delta \widehat H\widehat U(t)]\ , \ \ \ 
t={2\pi \hbar \over \Delta}\,{k\over N}=t_{\rm H}{k\over N}
\end{equation}
where $\hat U(t)=\exp (-{\rm i}\widehat Ht)$ is the evolution operator.
We consider first diffusive systems (electrons in disordered metals),
then systems with a chaotic classical limit.

{\em Diffusive systems}.
In this case we may consider the perturbation $\delta \widehat H$ to be
uncorrelated random changes of the site energies $V_{\bf n}$ in an 
Anderson tight-binding model \cite{cha96}
\begin{equation}
\label{eq:24}
\delta H=\sum_{\bf n} \delta V_{\bf n} \widehat P_{\bf n}\ , \ \ \ 
\langle \delta V_{\bf n}\delta V_{{\bf n}'}\rangle=\delta_{{\bf n},{\bf n}'}
\end{equation}
where $\widehat P_{\bf n}$ is the projection for locating an electron
on lattice site ${\bf n}$.
Using the semiclassical approximation (\ref{eq:23}),
$I(t)$ is seen to to be proportional to the probability
of returning to the original site after time $t$. Normalising
so that $I(t)$ approaches unity
for large $t$, we have
\begin{equation}
\label{eq:25}
I(t) = \sum_{\nu=0}^\infty {\rm e}^{-D k_\nu^2 t}
\end{equation}
where the sum is over the eigenmodes
of the Helmholtz equation
$(\nabla^2+k_\nu^2)\psi_\nu({\bf r}) = 0$
with Neumann boundary conditions.
In a quasi-one dimensional
system, $k_\nu = \pi\nu/L$.
In this case we obtain from Eq. (\ref{eq:21}) 
\begin{eqnarray}
\label{eq:26}
\label{eq:var3}
\sigma_\beta^2(n) &=& \frac{2\Delta^2}{\beta\pi^2} 
\Big [\log(2\pi n) \!+\! C_\beta
\!+\!\frac{1}{2} \sum_{\nu=1}^\infty
\log\!\big(1\!+\!\frac{n^2}{g^2\nu^4}\,\big)\Big]
\end{eqnarray}
where $g = \pi^2 \hbar D/L^2\Delta$ is a 
dimensionless conductance. 

\begin{figure}
\vspace*{-0.5cm}
\centerline{\psfig{file=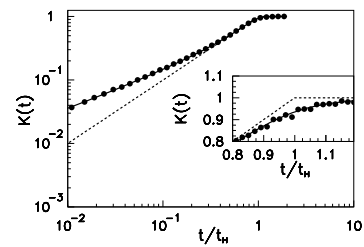,width=5cm}}
\caption{\label{fig:kt} The form factor $K(t)$
for a quasi-one dimensional diffusive
system. Numerical simulations for banded 
random matrices ($\bullet$), are compared with 
theoretical results ($-\!\!\!-\!\!\!-\!\!\!-$)
according to Eqs. (\protect\ref{eq:9}),
(\protect\ref{eq:10}) and (\protect\ref{eq:26}).
The GUE result is also shown
($---$).}
\end{figure}

An alternative way to express the results
for $R_\beta(\epsilon)$
is in terms of the \lq form factor' $K(t)$ which
is the Fourier transform of $R_\beta(\epsilon)$. As discussed
by AA, Eq. (\ref{eq:3}) gives rise
to non-universal structures in the form factor
at the Heisenberg time $t_{\rm H}$.
We have verified their existence
in numerical simulations: we used
an ensemble of complex Hermitian banded
random matrices (of dimension $N=1000$
and band width $b=35$) modelling
a quasi-one dimensional diffusive system.
We fitted the dimensionless conductance, using
states from the centre of the spectrum
(obtaining $g\simeq 2.0$).
The results are shown in Fig. \ref{fig:kt}
and are in good agreement with
the theoretical predictions, 
Eqs. (9), (10) and (26).

The theoretical predictions shown in Fig. 1
are precisely equivalent to the AA results
(7) and (14) in \cite{and95}.
We have thus shown that their results are 
consistent with the much simpler approach to 
justifying random matrix theory discussed in \cite{dys72}.

{\em Classically chaotic systems.}
Consider the expression (\ref{eq:23}) for the fluctuation
$\delta a_k$ of the Fourier coefficients. If the system
has a smooth classical Hamiltonian, this expression will clearly
be negligible unless $t$ corresponds approximately
with the period of a periodic orbit. Moreover, it was
shown in \cite{wil88} that when the motion is chaotic 
the $\delta a_k$ have statistics corresponding to random 
matrix theory for large $t$. A simple model 
capturing the essential features of this case is to
replace the lower limit of the integral in (\ref{eq:21}) with 
the period $t_0$ of the shortest periodic orbit.
This gives:
\begin{eqnarray}
\sigma^2_\beta(n)&\simeq&\frac{2\Delta^2}{\beta\pi^2}
\biggl[{\rm Ci}\big(\frac{2\pi n t_0}{t_{\rm H}}\big)\!-\!
\log\big(\frac{t_0}{t_{\rm H}}\big)\!-\!\gamma\!+\!C_\beta\biggr]\,.
\label{eq:27}
\end{eqnarray}
This model has the feature that $\sigma^2_\beta(n)$ is finite 
in the limit $n\to \infty$, corresponding to the 
behaviour of Dyson's $\Delta_3$ statistic
for systems with a smooth classical limit \cite{ber85}. 
We note that this
implies [using (\ref{eq:2}) and (\ref{eq:3})] that the 
oscillatory part of the correlation function does not
decay to zero. 
The form factor is seen
to have a delta function at the Heisenberg time $t_{\rm H}$,
with a magnitude proportional to $(t_0/t_{\rm H})^{4/\beta}$. 
This feature has not been remarked upon in earlier papers
which have discussed the form factor for chaotic systems
\cite{aga95,bog96,pra96}.

A more precise estimate of $\sigma^2_\beta(n)$ for specific 
systems can be obtained using periodic orbit theory, following
the approach used in \cite{ber85}. The conclusions are 
unchanged: $\sigma^2_\beta(n)$ remains finite as $n\to \infty$,
and there must exist oscillations in the correlation function
which do not decay. One difference is that for large $n$ 
the $p_\beta(n,\epsilon)$ are dominated by 
the shortest orbits. Because only a finite number of 
components are significant, the central limit theorem cannot
be used to assert that these distributions are Gaussian.
Moreover small deviations from a Gaussian distribution
can have a large effect on the Fourier transform of the
distribution, which determines the magnitude of the oscillations
in (\ref{eq:10}). We infer that the AA relations
may need to be modified when applied to the spectra of systems 
with a smooth classical limit.

{\em Conclusions}.
We have shown that the AA relations have a simple interpretation, 
independent of random matrix theory and supersymmetric techniques.
We have shown how spectral correlations of diffusive systems
are obtained using from Dyson's Brownian motion model,
indicating that the AA relation is consistent with this
approach.

{\em Acknowledgements}.
It is a pleasure to acknowledge illuminating discussions
with O. Agam.
BM was partially supported by the SFB 393.
MW was supported by the Max Planck Institute for
the Physics of Complex Systems, Dresden, and the EPSRC, grant GR/L02302.

\narrowtext

\end{multicols}
\end{document}